
%
%
\input phyzzx

%
\catcode`\@=11 
\def\papersize{\hsize=40pc \vsize=53pc \hoffset=0pc \voffset=1pc
   \advance\hoffset by\HOFFSET \advance\voffset by\VOFFSET
   \pagebottomfiller=0pc
   \skip\footins=\bigskipamount \normalspace }
\catcode`\@=12 
\papers
\def\to{\rightarrow}

\vsize=23.cm
\hsize=15.cm

\tolerance=500000
\overfullrule=0pt

\Pubnum={LPTENS-95/32 \cr
{\tt hep-th@xxx/9508062} \cr
August 1995}

\date={}
\pubtype={}
\titlepage
\title{\bf Consistent string backgrounds and completely integrable 2D field
theories}
\author{Adel~Bilal}
\address{
CNRS - Laboratoire de Physique Th\'eorique de l'Ecole
Normale Sup\'erieure
\foot{{\rm unit\'e propre du CNRS, associ\'e \`a l'Ecole Normale
Sup\'erieure et l'Universit\'e Paris-Sud}}
 \nextline 24 rue Lhomond, 75231
Paris Cedex 05, France\break
{\tt bilal@physique.ens.fr}
}

\abstract{After reviewing the $\beta$-function equations for consistent string
backgrounds in the $\sigma$-model approach, including metric and antisymmetric
tensor,
dilaton and tachyon potential, we apply this formalism to WZW models. We
particularly
emphasize the case where the WZW model is perturbed by an integrable marginal
tachyon
potential operator leading to the non-abelian Toda theories. Already in the
simplest such
theory, there is a large non-linear and non-local chiral algebra that extends
the
Virasoro algebra. This theory is shown to have two formulations, one being a
classical
reduction of the other. Only the non-reduced theory is shown to satisfy the
$\beta$-function equations.}

\vskip 1.cm
\centerline{\it To appear in the Proceedings of the Trieste Conference on }
\centerline{\it Recent Developments in Statistical Mechanics and Quantum Field
Theory, April 1995}

\endpage
\pagenumber=1

 \def\PL #1 #2 #3 {Phys.~Lett.~{\bf #1} (#2) #3}
 \def\NP #1 #2 #3 {Nucl.~Phys.~{\bf #1} (#2) #3}
 \def\PR #1 #2 #3 {Phys.~Rev.~{\bf #1} (#2) #3}
 \def\PRL #1 #2 #3 {Phys.~Rev.~Lett.~{\bf #1} (#2) #3}
 \def\CMP #1 #2 #3 {Comm.~Math.~Phys.~{\bf #1} (#2) #3}
 \def\IJMP #1 #2 #3 {Int.~J.~Mod.~Phys.~{\bf #1} (#2) #3}
 \def\JETP #1 #2 #3 {Sov.~Phys.~JETP.~{\bf #1} (#2) #3}
 \def\PRS #1 #2 #3 {Proc.~Roy.~Soc.~{\bf #1} (#2) #3}
 \def\IM #1 #2 #3 {Inv.~Math.~{\bf #1} (#2) #3}
 \def\JFA #1 #2 #3 {J.~Funkt.~Anal.~{\bf #1} (#2) #3}
 \def\LMP #1 #2 #3 {Lett.~Math.~Phys.~{\bf #1} (#2) #3}
 \def\IJMP #1 #2 #3 {Int.~J.~Mod.~Phys.~{\bf #1} (#2) #3}
 \def\FAA #1 #2 #3 {Funct.~Anal.~Appl.~{\bf #1} (#2) #3}
 \def\AP #1 #2 #3 {Ann.~Phys.~{\bf #1} (#2) #3}
 \def\MPL #1 #2 #3 {Mod.~Phys.~Lett.~{\bf #1} (#2) #3}

\def\d{\partial}

\def\vf{\varphi}
\def\ix{\int {\rm d}^2 x \, }
\def\e{\epsilon}

\def\a{\alpha}
\def\b{\beta}
\def\m{\mu}
\def\n{\nu}
\def\r{\rho}
\def\l{\lambda}
\def\g{\gamma}

\def\xp{x^+}

\def\rd{\sqrt{2}}

\def\dd #1 #2{{\delta #1\over \delta #2}}
\def\tr{{\rm tr}\ }

\def\rg{\sqrt{g}}

\def\G{{\cal G}}
\def\b{\beta}
\def\s{\sigma}
\def\ap{\a'}
\def\ids{\int {\rm d}^2\s}
\def\O{{\cal O}}
\def\R{{\cal R}}
\def\tr{{\rm tr}}
\def\x{\xi}
\def\o{\omega}
\def\cg{{\sl g}}
\def\ch{{\rm ch}}
\def\sh{{\rm sh}}
\def\th{{\rm th}}
\def\ds{\delta(\s-\s')}
\def\es{\e(\s-\s')}

\REF\CAL{C.G. Callan, D. Friedan, E.J. Martinec and M.J. Perry, \NP
B262 1985 593 .}
\REF\POLY{A. Polyakov, \PL B103 1981 207 ;\nextline
J.-L. Gervais and A. Neveu,
\NP B209 1982 125 ;\nextline
E. Braaten, T.L. Curtright, C.B. Thorn, \PL 118B
1982 115 .}
\REF\BGT{A. Bilal and J.-L. Gervais, \PL B206 1988 412 ;
\NP B314 1989 646 , {\bf B318} (1989) 579 .}
\REF\LIOU{J. Liouville, Comptes Rendus Acad. Sc. (Paris), vol. XXXVI (1883)
71.}
\REF\LS{A.N. Leznov and M.V. Saveliev, \CMP 89 1983 59 .}
\REF\GERBAK{J.-L. Gervais, \PL 160B 1985 277 ;\nextline
I. Bakas, \CMP 123 1989 627 .}
\REF\GS{J.-L. Gervais and M.V. Saveliev, \PL B286 1992 271 .}
\REF\NAT{A. Bilal, \NP B422 1994 258 .}
\REF\DVV{R. Dijkgraaf, E. Verlinde and H. Verlinde, \CMP 115 1988 649 .}
\REF\KS{I. Klebanov and L. Susskind, \PL B200 1988 446 ;\nextline
S.R. Das and B. Sathiapalan, \PRL 56 1986 2664 ;\nextline
E.S. Fradkin and A.A. Tseytlin, \NP B261 1985 1 .}
\REF\ZACH{E. Braaten, T.L. Curtright and C.K. Zachos, \NP B260 1985 630 .}
\REF\LMP{A. Bilal, Lett. Math. Phys. {\bf 32} (1994) 103.}
\REF\MW{A. Bilal, \CMP 170 1995 117 .}
\REF\GSWP{J.-L. Gervais and M.V. Saveliev, work in progress.}
\REF\BGWS{A. Bilal and J.-L. Gervais, \NP B326 1989 222 .}

\chapter{Introduction}

Consistent string backgrounds are given by exact conformal field
theories. If the string  propagating on a curved background is
described by the corresponding two-dimensional (non-linear)
$\s$-model action, conformal invariance is equivalent to the
vanishing of the associated $\b$-functionals
[\CAL]. Part of the string background (``internal dimensions") can of
course be replaced by a more general exact conformal field theory.
Particular examples are actions that involve exactly marginal
operators added to a free bosonic action, as is the case for the
Liouville
[\POLY] and Toda
[\BGT] actions. The latter theories are highly non-trivial, but
nevertheless are known to be completely integrable since long
[\LIOU,\LS].
They are intimately related to the Gelfand-Dikii hierarchies of
integrable partial differential equations generalizing the KdV
equation
[\GERBAK].

In this note, I will discuss certain integrable theories that again
lead to consistent string backgrounds, i.e. are exactly conformal.
 From the algebraic point of view, they are associated with
non-canonical gradations of simple Lie algebras $\cg$ where the
gradation-zero part $\cg_0$ is a {\it non-abelian} subalgebra. The prototype is
the non-abelian Toda theory studied in
[\GS,\NAT].
An important point to be made is that in the formulation of [\GS,\NAT]
the non-abelian Toda theory does not satisfy the $\b$-function equations of
[\CAL]. However, as will be shown below, there exists a classically equivalent,
and more natural formulation that does.

This note is organized as follows: In section 2, after briefly reviewing the
setting
of [\CAL] for the $\b$-function equations in the $\s$-model, we discuss the
role of
the tachyon potential. In section 3, we show how the WZW-models naturally fit
into
this setting. In sect. 4, we introduce the non-abelian Toda theory in both
classically
equivalent formulations and show how the  $\b$-function equations are satisfied
in one
of these formulations. We point out that this should actually be no surprise,
at
least as far as the equations for the metric, antisymmetric tensor and dilation
are concerned, since they are given by those of a WZW-model for $\cg_0$.

\chapter{Consistent string backgrounds in the $\s$-model}

Consider string propagation on an arbitrary background described by the
(non-linear)
$\s$-model
$$\eqalign{
S_\s=&-{1\over 4\pi\ap} \ids \rg \left[ G_{\m\n}(X)g^{\a\b}\d_\a X^\m \d_\b
X^\n
+B_{\m\n}(X)\e^{\a\b}\d_\a X^\m \d_\b X^\n \right]  \cr
&+{1\over 4\pi} \ids \rg R^{(2)} \Phi(X) \cr}
\eqn\i$$
where $G_{\m\n}, B_{\m\n}$ and $\Phi$ are the background metric, antisymmetric
tensor
and dilaton field. $\e_{\a\b}$ and $g_{\a\b}$ are the world-sheet antisymmetric
$\e$-tensor and metric (with signature $(-,+)$), and $R^{(2)}$ is the
corresponding world-sheet curvature scalar. The string tension $\ap$ plays the
role of a loop-counting parameter ($\sim\hbar$) for the first part of the
action. The dilaton-term, being a sort of anomaly cancellation term, is
$\O((\ap)^0)$.

Callan et al have shown [\CAL] that, to lowest non-trivial order in $\ap$, the
so-defined string theory is conformally invariant, i.e
 $G_{\m\n}, B_{\m\n}$ and $\Phi$ are consistent string backgrounds, if the
following
$\b$-function equations are satisfied:
$$\eqalign{
\b^{G_{\m\n}}
&\sim \R_{\m\n}-{1\over 4}H_{\m\r\s}H_\n^{\phantom{\n}\r\s} +2D_\m D_\n \Phi =0
\ , \cr
\b^{B_{\m\n}}
&\sim D_\r H^\r_{\phantom{\r}\m\n} -2 (D_\r\Phi)H^\r_{\phantom{\r}\m\n} =0\ ,
\cr
\b^\Phi
&\sim {N-26\over 3\ap} +4D_\m\Phi D^\m\Phi - 4 D_\m D^\m \Phi - \R
+{1\over 12} H_{\m\n\r}H^{\m\n\r} =0 \ .\cr}
\eqn\ii$$
Here $H_{\m\n\r}=\d_\m B_{\n\r}+\d_\n B_{\r\m} +\d_\r B_{\m\n}$, and
$\R_{\m\n}$
and $\R$ are the Ricci curvature tensor and scalar computed
\foot{
Our sign convention is $\R_{\m\n}=R^\l_{\phantom{\l}\m\l\n}$,
$R^\m_{\phantom{\m}\n\r\s}=(\d_\r\Gamma_\s-\d_\s\Gamma_\r
+[\Gamma_\r,\Gamma_\s])^\m_{\phantom{\m}\n}$,
$(\Gamma_\r)^\m_{\phantom{\m}\n}=\Gamma^\m_{\r\n}={1\over 2} G^{\m\l}\left(
\d_\r
G_{\l\n}+\d_\n G_{\r\l}-\d_\l G_{\r\n}\right)$.
}
 from the space-time metric
$G_{\m\n}$, while $D_\m$ is the corresponding covariant derivative. $N$ is the
number
of fields $X^\m$.

In many cases one wishes to study the effect of adding a perturbation to the
conformal
theory. In particular one can ask when this perturbation is exactly marginal so
that
the resulting theory remains an exact conformal field theory. If the
perturbation
simply modifies  $G_{\m\n}, B_{\m\n}$ and $\Phi$ one just needs to check
whether the
new background fields still satisfy eqs. \ii. But there are many other
operators one
can add, the simplest of which is the so-called tachyon potential, i.e. a
non-derivative function of the fields $X^\m$. We could choose to add to the
action \i\
either a term
$-{1\over 4\pi\ap} \ids \rg \, {{\tilde V(X)}}$ or, working in conformal gauge
 right away, a term
$$S_V=-{1\over 4\pi\ap} \ids\, V(X)\ .
\eqn\iia$$
While ${{\tilde V(X)}}$ should be a scalar, conformal field theory tells us
that,
to be marginal, $V(X)$ should be a conformal primary of weight $(1,1)$.
\foot{
We also know that an integrable marginal operator moreover needs  to have
vanishing operator product coefficients $c_{VVj}$ with all primary fields
labelled by $j$ that have weight $(1,1)$ [\DVV].
}

In the $\s$-model approach one can again compute the $\b$-functionals now
including
the effect of $V$. To lowest order in $\ap$ one finds that eqs \ii\ are not
modified
by the presence of the tachyon potential $V(X)$, while $\b^V$ is
$$\b^V \sim -{1\over 2} D^\m D_\m V + D^\m \Phi D_\m V - {1\over \ap} V \ .
\eqn\iib$$
We insisted that these results are to lowest non-trivial order in $\ap$.
Indeed, it is
known
[\KS]
that including higher-order contributions in $\ap$ will lead to corrections to
eqs
\ii\ that involve $V$, as well as to non-linear terms in $V$ in eq \iib. As it
stands, eq. \iib\
is simply the dimension $(1,1)$-condition in the $\s$-model language. Typical
solutions of $\b^V=0$ are sums of products of exponentials of the fields.

As an example consider the Liouville theory with $G_{\vf\vf}=\ap$, $\Phi=Q\vf$
and $V\sim e^{\b\vf}$, coupled to a matter theory with central charge $c$.
Substituting $N=1+c$ in the third eq. \ii\ yields $Q^2={25-c\over 12}$.
Equation \iib\ gives $\b^2-2Q\b+2=0$, hence the well-known formula
$\b=Q-\sqrt{Q^2-2}$ $=\sqrt{{25-c\over 12}}-\sqrt{{1-c\over
12}}$. In the classical limit $c\to -\infty$ one correctly gets $\b\to
{1\over Q}$. Upon rescaling $\tilde\vf={\vf\over Q} + const$, the Liouville
action correctly
reads [\POLY] ${25-c\over 48\pi} \ids \left[ -(\d\tilde\vf)^2 +R\tilde\vf
- e^{(\b Q)\tilde \vf} \right]$ with $\b Q \sim 1$ in this limit.

Of course, even in the absence of the potential $V$, eqs \ii\ are valid only to
leading order in $\ap$ and higher order corrections e.g. to the first equation
involve terms like $R_{\m\l\r\s}R_{\n}^{\phantom{\n}\l\r\s}$. In some cases,
however, as for the WZW-models to which we now turn, one knows that the theory
is an exact conformal theory, and hence the $\b$-functionals vanish to all
orders in $\ap$.

\chapter{WZW-models}

It is very easy to compute the background fields $G_{\m\n}$ and $H_{\m\n\r}$
for an
arbitrary WZW-model  for a group $\G$
[\ZACH]
 with action
$$\eqalign{
S=S_1+S_2=&{1\over 16\pi\ap} \ids\, \tr (g^{-1}\d_\a g) (g^{-1}\d^\a g) \cr
& +{1\over 24 \pi\ap} \ids {\rm d}t\, \e^{\a\b\g}\, \tr (g^{-1}\d_\a g)
(g^{-1}\d_\b g) (g^{-1}\d_\g g)  \ . \cr }
\eqn\iii$$
Let $\x^\m$ be coordinates on the group manifold of $\G$ so that $g\equiv
g(\x^\m(\s^\a))$. Let $\d_\m \equiv {\d\over \d\x^\m}$. Then $(\d_\m g) g^{-1}$
is in
the Lie algebra of $\G$ and hence some linear combination of the generators
$I_a$:
$$\d_\m g =U^a_\m(\x) I_a g
\eqn\iv$$
and from equating $\d_\m\d_\n g$ with $\d_\n\d_\m g$ one obtains the
Maurer-Cartan
equations
$$\d_\m U^a_\n -\d_\n U^a_\m=C^a_{\phantom{a}bc} U^b_\m U^c_\n
\eqn\v$$
where $[I_b,I_c]=C^a_{\phantom{a}bc} I_a$. We normalize the generators as $\tr
I_aI_b=-\delta_{ab}$. Furthermore, $C_{acd} C_b^{\phantom{b}cd}=c_2
\delta_{ab}$ where
$c_2$ is the quadratic Casimir of the adjoint representation.

Using eq. \iv\ it is straightforward to see that the first part of the action
reduces to
$$S_1=-{1\over 4\pi\ap} \ids\, \d_\a\x^\m \d^\a \x^\n G_{\m\n}
\eqn\vi$$
with
$$G_{\m\n}={1\over 4} U^a_\m U^a_\n \ .
\eqn\vii$$
Thus we interpret the $e^a_\m = {1\over 2}U^a_\m$ as vielbeins on the group
manifold.
Then eq. \v\ is nothing but the zero-torsion equation ${\rm d}
e^a+\o^a_{\phantom{a}b}\wedge e^b=0$, provided we identify the spin-connection
as
$$\o^{\phantom{\mu}a}_{\m\phantom{a}b}={1\over 2} C^a_{\phantom{a}bc} U^c_\m\ .
\eqn\viii$$
It is an easy excercise to compute the curvature two-form $R^a_{\phantom{a}b}=
{\rm d}\o^a_{\phantom{a}b}+ \o^a_{\phantom{a}c}\wedge \o^c_{\phantom{c}b}$ and
the Riemann tensor $R_{\m\n\r\s}=e_{a\r}e^b_\s  (R^a_{\phantom{a}b})_{\m\n}$
from
which  the Ricci tensor and scalar are obtained as
$$\R_{\m\n}=c_2 G_{\m\n}\ , \quad \R=c_2\, {\rm dim}\G \ .
\eqn\ix$$
The Ricci tensor is proportional to the metric and the scalar curvature is
constant.
\foot{
This is reminiscent of a maximally symmetric space. However, only for $SU(2)$
where
$\G \simeq S^3$ one has the extra relation $R_{\m\n\r\s}\sim
G_{\m\r}G_{\n\s}-G_{\m\s}G_{\n\r}$.
}

While the first part $S_1$ of the WZW-action gives the metric $G_{\m\n}$,
the WZ-term $S_2$ will give the antisymmetric tensor field. It is not possible
in
general to give $B_{\m\n}$ directly (otherwise we would have obtained a
two-dimensional form of the WZ-term), but it is easy to obtain its field
strength
$H_{\m\n\r}$ which is all that is needed anyway. Using eq. \iv\ we get
$$S_2=-{1\over 48 \pi\ap} \ids {\rm d}t\,
\e^{\a\b\g}\d_\a\x^\m\d_\b\x^\n\d_\g\x^\r C_{abc}U^a_\m U^b_\n U^c_\r
\eqn\xp$$
which we want to identify with
$$\eqalign{
&-{1\over 4 \pi\ap} \ids\,  \e^{\a\b} \d_\a\x^\m\d_\b\x^\n B_{\m\n}
=-{1\over 4 \pi\ap} \ids  {\rm d}t\, \e^{\a\b\g} \d_\g ( \d_\a\x^\m\d_\b\x^\n
B_{\m\n}) \cr
&=-{1\over 4 \pi\ap} \ids  {\rm d}t\, \e^{\a\b\g}
\d_\a\x^\m\d_\b\x^\n\d_\g\x^\r
\d_\r B_{\m\n} \cr }
\eqn\xpi$$
(where, as usual, the boundary of the three-dimensional integration region is
the
two-dimensional manifold of the $\s$-model) so that
$$H_{\m\n\r}= {1\over 4} C_{abc} U^a_\m U^b_\n U^c_\r
=2 C_{abc}e^a_\m e^b_\n e^c_\r \ .
\eqn\xii$$

Since the vielbeins $e^a_\m$, as well as the structure constants $C_{abc}$ are
covariantly constant, it follows that $D_\s H_{\m\n\r}=0$. Furthermore
$H_{\m\r\s}H_\n^{\phantom{\n}\r\s}=4 c_2 G_{\m\n}=4\R_{\m\n}$, so that the
three eqs. \ii\ reduce to
$$\eqalign{
G^{\m\n}D_\m\d_\n \Phi &=0 \ , \cr
\d_\r \Phi H^\r_{\phantom{\r}\m\n}  &=0 \ , \cr
G^{\m\n}D_\m \Phi D_\n \Phi +{\tilde N-26\over 12\ap}&= -{{\rm dim}\G\over
12\ap}(1-2\ap
c_2) \ , \cr }
\eqn\xiii$$
which determines $\Phi$. Here $\tilde N$ is the number  of fields other than
the
$\x^\m$, if any.
\foot{
Note that the r.h.s. of the last equation is, to order $\ap$ and
up to an overall factor of $-{1\over 12\ap}$, the contribution of the WZW-model
to the
central charge: $c={x{\rm dim}\G\over x+\tilde h}={\rm dim}\G (1-2\ap c_2)
+\O({\ap}^2)$
where $\tilde h=c_2/\psi^2$ ($\psi^2$ being the length  squared of the highest
root) is the dual Coxeter number, and one identifies the Kac-Moody level $x$
with
${1\over 2\ap\psi^2}$
{}.
}
For semi-simple $\G$, the $C_{abc}$ cannot vanish for all $a,b$ for fixed $c$,
hence
$ H^\r_{\phantom{\r}\m\n} $ cannot vanish for all $\m,\n$ for fixed $\r$, and
one concludes
that $\d_\r \Phi=0$. Then, for $\tilde N=0$, unless one fine-tunes $\ap$, i.e.
unless the
WZW Kac-Moody level can be adjusted such that $c_{\rm WZW}=26$, the third eq.
\xiii\ is not
satisfied.

If, however, we consider the slightly more general situation where in addition
to
the WZW fields $\x^\m$ one has $\tilde N$ other fields $\x^s,\ s={\rm dim }\G
+1, \ldots  {\rm dim }\G +\tilde N$ with flat metric and vanishing $B_{\m\n}$
($U(1)$-factors) then, in these directions $H^s_{\phantom{s}\m\n}=0$ and
$\d_s\Phi$ can be non-vanishing. The first eq. \xiii\ then shows that the
dilaton field must be linear in the $\x^s$ and one has %
$$\eqalign{
\Phi&=\sum_{ s={\rm dim }\G +1}^{{\rm dim }\G +\tilde N} a_s \x^s \cr
\sum_{ s={\rm dim }\G +1}^{{\rm dim }\G +\tilde N} a_s^2 &= {26-\tilde N -{\rm
dim}\G\over
12\ap} +{c_2\, {\rm dim }\G\over 6} \cr }
\eqn\xiv$$
so that all $\b$-function equations \ii\ are satisfied.

\chapter{The non-abelian Toda theory}

The (ordinary) conformally invariant Toda field theories based on a  Lie
algebra
$\cg$
[\BGT]
when viewed as a $\s$-model \i\ have constant metric $G_{\m\n}$ and vanishing
antisymmetric
tensor, and thus the corresponding $\b$-function equations \ii\ are trivially
satisfied,
provided one chooses a linear dilaton appropriately. The latter is invisible in
the
conformal gauge action but controls the improvement term in the stress-energy
tensor. These
theories correspond to a canonical gradation of $\cg$ and the gradation zero
part $\cg_0$
is abelian. Generalizing to non-canonical gradations leads to non-abelian
$\cg_0$
[\LS]
and non-trivial background fields
[\GS].
The simplest example is the non-abelian Toda theory for $\cg = B_2$. In the
formulation
given in
[\GS,\NAT]
it corresponds to a $\s$-model with the metric of the two-dimensional black
hole
and one additional flat dimension:
$$G_{\m\n}{\rm d}x^\m {\rm d}x^\n= {\rm d}r^2 +\th^2r\, {\rm d}t^2 +{\rm
d}\phi^2\ , \quad B_{\m\n}=0
\eqn\xv$$
and a potential term $V=4 \ch 2r e^{2\phi}$.
This model is classically integrable and although its equations of motion are
highly
non-trivial, their general solution could be explicitly given
[\GS,\NAT]
due to the underlying Lie algebraic structure. At the classical (Poisson
bracket) level
this model has three left-moving conserved quantities $T, V^+, V^-$ of
dimension two, as
well as three right-moving ones. The chiral equal-time Poisson bracket algebra
of these
quantities is not only non-linear (like for $W$-algebras) but also {\it
non-local} due to
the appearance of $\e(\s-\s')=\theta(\s-\s')-\theta(\s'-\s)$:
\def\gmd{\gamma^{-2}}

\def\dsppp{\delta'''(\s-\s')}
$$\eqalign{
\gmd \{T(\s)\, ,\, T(\s')\}  &=
(\d_\s-\d_{\s'})\left[ T(\s') \ds\right]-{1\over 2} \dsppp \cr
\gmd \{T(\s)\, ,\, V^\pm(\s')\}  &=
(\d_\s-\d_{\s'})\left[ V^\pm(\s') \ds\right]\cr
\gmd \{V^\pm(\s)\, ,\, V^\pm(\s')\}&=\es
V^\pm(\s)V^\pm(\s')\cr
\gmd \{V^\pm(\s)\, ,\, V^\mp(\s')\}&=-\es
V^\pm(\s)V^\mp(\s')\cr
&\phantom{=}+(\d_\s-\d_{\s'})\left[ T(\s') \ds\right] -{1\over 2}
\dsppp \ .\cr }
\eqn\xvi$$
The scale factor $\g^2$ which controls the central charge plays the role of
$\hbar$ or $\ap$ already introduced in the classical equations. The classical
solution of the non-abelian Toda theory provides a free-field realisation of
these chiral generators [\NAT].
It turned out that this algebra can be obtained as the second Gelfand-Dikii
symplectic
structure associated with a second-order matrix differential operator
[\LMP]
$$ L=\d^2-U\ ,\quad
U=\pmatrix{ T& -\rd V^+\cr -\rd V^- & T \cr }
\eqn\xvii$$
in the same way as the Virasoro algebra is obtained from  \xvii\ with scalar
$U\sim T$. This algebraic structure was generalized to
$n\times n$-matrix  differential operators of order $m$ and non-linear,
non-local algebras
like \xvi\ that are matrix generalizations of (classical) $W_m$-algebras
[\MW].

All this was at the level of classical symplectic structures, and it seemed
surprisingly
difficult to quantize even the simplest algebra \xvi\ maintaining the conformal
dimensions
of the generators equal to two. It is probable that the solution of the
quantization problem is linked to the following simple remark.

It is easy to see that the black hole metric \xv\ cannot solve the
$\b$-function equations
\ii, i.e. one cannot find a dilation $\Phi$ so that all three equations are
satisfied: the
non-abelian Toda model so defined is classically conformally invariant, but not
at the
quantum level. The point is that the $\s$-model action corresponding to \xv\ is
{\it not}
the correct starting point.

In the original Lax pair formulation of the integrable model there are four
fields $\m, \n,
r, \phi$ whose equations of motion can be written as
[\GSWP]
$$\eqalign{
&\d_-(\d_+\n-\ch 2r \d_+\m) = 0 \ , \cr
&\d_+(\d_-\m-\ch 2r \d_-\n) = 0 \ , \cr
&2\d_+\d_- r = \sh 2r (\d_-\n\d_+\m +2e^{2\phi}) \ , \cr
&\d_+\d_-\phi = \ch 2r e^{2\phi} \ , \cr }
\eqn\xviii$$
which can be obtained from the action
$$\eqalign{
S\sim \ids \Big[ &-{1\over 2} \d_+\n\d_-\n   -{1\over 2} \d_+\m\d_-\m +2 \d_+
r\d_- r
+2\d_+\phi\d_-\phi \cr
&+\ch 2r \d_+\m \d_-\n +2\ch 2r\, e^{2\phi} \Big] \ . \cr }
\eqn\xix$$
We can solve the first two equations \xviii\ as
$$\d_+\n-\ch 2r\, \d_+\m = 0 \ , \quad \d_-\m-\ch 2r\, \d_-\n=0\ .
\eqn\xx$$
Introduce a field $t$ as
$$\d_+ t = \ch^2 r\, \d_+\m \ , \quad \d_- t=(1+\th^2 r )^{-1} \d_-\m \ .
\eqn\xxi$$
Then eq. \xx\ implies that $\d_+\n=(1+\th^2 r ) \d_+ t\ , \ \d_-\n= \ch^{-2}
r\,
\d_- t$. The compatibility condition of the two equations \xxi\ can be written
as
$$\d_+\d_- t =-{1\over \sh r\, \ch r} (\d_+ t \d_- r + \d_- t \d_+ r)
\eqn\xxii$$
while the last two equations \xviii\ become
$$\eqalign{
\d_+\d_- r &={\sh r\over \ch^3r} \d_+ t\d_- t + \sh 2r\, e^{2\phi} \ , \cr
\d_+\d_- \phi &= \ch 2r\, e^{2\phi} \ . \cr }
\eqn\xxiii$$
Equations \xxii\ and \xxiii\ are three equations of motion for the three fields
$r, t,
\phi$ and can be obtained from the action
$$S\sim \ids \left[ \d_+ r\d_- r +\th^2 r\, \d_+ t \d_- t +\d_+\phi \d_-\phi
+\ch 2r\, e^{2\phi} \right]
\eqn\xxiv$$
which is the $\s$-model with the background fields given by \xv\ and a tachyon
potential
$V=4\ch 2r\, e^{2\phi}$, i.e. the non-abelian Toda action as formulated in
[\GS,\NAT].

While, as already
mentioned, the $\s$-model \xxiv\ does not satisfy the $\b$-function equations,
the
model given by \xix\ does, as we now poceed to show. Arranging the fields as
$X^1=\m,
X^2=\n, X^3=r, X^4=\phi$,  the action \xix\ corresponds to
$$G_{\m\n}={1\over 4}
\pmatrix{-1&\ch 2r &0&0\cr
\ch 2r&-1&0&0&\cr
0&0&4&0&\cr
0&0&0&4\cr} \quad , \quad
B_{\m\n}={1\over 4}
\pmatrix{0&-\ch 2r &0&0\cr
\ch 2r&0&0&0&\cr
0&0&0&0&\cr
0&0&0&0\cr}\ ,
\eqn\xxv$$
and $V=4\ch 2r\, e^{2\phi}$ as before. As far as $G_{\m\n}$ and $B_{\m\n}$ are
concerned, $\phi$ plays a trivial role, and it is more convenient to consider
seperately $X^1, X^2$ and $X^3$ with metric $G^{(3)}_{\m\n}$. It is then
straightforward to see that %
$$\R^{(3)}_{\m\n} = -2 G^{(3)}_{\m\n}
\eqn\xxvi$$
while the full four-dimensional Ricci tensor $\R_{\m\n}$
is obtained by adding an extra lign and column
of zeros. Relation \xxvi\ which closely ressembles eq. \ix\ for $SU(2)$ should
be no
surprise. Indeed, the $\s$-model part of the action \xix\ can be written as a
WZW-model for the zero-gradation subgroup $\G_0$ which is a non-compact version
of $SU(2)$, hence the minus sign. Similarly one finds $H_{123}=-{1\over 2}\sh
2r$ and $H_{\m\r\s}H_\n^{\phantom{\n}\r\s}=4\R_{\m\n}$, as well as $D_\r
H^\r_{\phantom{\r}\m\n}=0$ and the $\b$-function equations \ii\ again reduce to
eqs. \xiii\ with ${\rm dim}\G=3$ and $c_2=-2$. As in section 3 one determines
$$\Phi= Q\phi\ , \quad Q^2={22-d\over 12\ap}-1
\eqn\xxvii$$
where $d$ is the number of free (flat) extra fields one might want to add. Note
that
because $\R=-6=const$, the action \xix\ certainly no longer describes a black
hole
background.

We have seen that the ``kinetic" part of the ``correct" non-abelian Toda action
\xix\
satisfies the $\b$-function equations \ii. What about the potential
$V=4\ch 2r\, e^{2\phi}\, $? First, we remark that the normalization factor $4$
in
front of $e^{2\phi}$ has no meaning since it can be adjusted at will by
shifting
$\phi\to\phi+const$. Also the action \xix\ was obtained from purely classical
considerations and we could just as well have multiplied it by
some scale factor $\gmd$. Redefining all fields $X^\m\to \g X^\m$ would
absorb this factor while changing $\gmd V\to 4 \gmd \ch 2\g r\, e^{2\g \phi}$,
so
one might just as well consider
$$V=\ch 2\g r\, e^{2\g\phi}\ .
\eqn\xxviii$$
This is analogous to the  rescaling of the Liouville field discussed in section
2. Of course, in the quantum theory the value of $\g$ is no longer
arbitrary. Inserting $V$ into eq. \iib\ one finds that it indeed is a solution
of
$\b^V=0$ provided $\g$ is chosen to satisfy the resulting
algebraic equation with solution
$$
4\g = Q \pm \sqrt{Q^2-{4\over \ap}} \ .
\eqn\xxix$$
For all non-negative values of $d$ this leads to a complex value of $\g$, and
it seems that
in order to quantize this theory one faces a situation similar to the Liouville
theory for
$c>1$.

Whether this is really so is not clear at present. Indeed, the world-sheet
symmetry algebra
is not just the Virasoro algebra which gives a ghost contribution $-26$ in
$\b^V$ and hence
in $-Q^2={d-4-26\over 12\ap}+1$, but the larger algebra \xvi. From the study of
superstrings or $W$-gravity and $W$-strings
[\BGWS]
one knows that the ghost contributions to the central charge gets modified, and
for a
larger bosonic chiral algebra it becomes larger (e.g. $-100$ instead of $-26$
for the
$W_3$-case). For the algebra \xvi\ one would then expect three ghost pairs, all
of weight 2
and $-1$, contributing $3\times (-26)=-78$ to the central charge, so that the
correct value
of $Q$ should read
$$Q^2={78-4-d\over 12\ap}-1 \ ,
\eqn\xxx$$
and $Q$ and $\sqrt{Q^2-{4\over\ap}}$ remain real as long as $d\le 26-12\ap$.
Curiously
enough, for $\ap\to 0$ the upper limit is $26$. However, the whole issue of
quantization of
the theory needs to be studied in detail before one can draw any definit
conclusion.

\ack

I am grateful to J.-L. Gervais and M.V. Saveliev for sharing their insights on
the
non-abelian Toda theories, and in particular for drawing my attention to eqs.
\xviii\ and
\xix.

\refout

\end